\documentclass[aps,prl,showpacs,twocolumn,superscriptaddress]{revtex4}
\usepackage{graphicx}
\usepackage{amssymb}
\usepackage{epsfig}
\usepackage{graphicx}
\usepackage{amsmath}
\usepackage{array,color}
\usepackage{leftidx}
\usepackage{mathrsfs}
\usepackage{ulem}
\usepackage{color}
\usepackage{float}
\usepackage{hyperref}

\begin{document}
\title{Exotic Vortex States with Discrete Rotational Symmetry in Atomic Fermi Gases with Spin-Orbital-Angular-Momentum Coupling}
\author{Liang-Liang Wang}
\affiliation{School of Science, Westlake University, 18 Shilongshan Road, Hangzhou 310024, Zhejiang Province, China}
\affiliation{Institute of Natural Sciences, Westlake Institute for Advanced Study, 18 Shilongshan Road, Hangzhou 310024, Zhejiang Province, China}
\author{An-Chun Ji}
\affiliation{Department of Physics,
Capital Normal University, Beijing 100048, China}
\author{Qing Sun}
\email{sunqing@cnu.edu.cn}
\affiliation{Department of Physics,
Capital Normal University, Beijing 100048, China}
\author{Jian Li}
\email{lijian@westlake.edu.cn}
\affiliation{School of Science, Westlake University, 18 Shilongshan Road, Hangzhou 310024, Zhejiang Province, China}
\affiliation{Institute of Natural Sciences, Westlake Institute for Advanced Study, 18 Shilongshan Road, Hangzhou 310024, Zhejiang Province, China}

\date{{\small \today}}

\begin{abstract}
We investigate the superfluidity of a two-component Fermi gas with spin-orbital-angular-momentum coupling (SOAMC). Due to the intricate interplay of SOAMC, two-photon detuning and atom-atom interaction, a family of vortex ground states emerge in a broad parameter regime of the phase diagram, in contrast to the usual case where an external rotation or magnetic field is generally required.
More strikingly, an unprecedented vortex state, which breaks the continuous rotational symmetry to a discrete one spontaneously, is predicted to occur. The underlying physics are elucidated and verified by numerical simulations. The unique density distributions of the predicted vortex states enable a direct observation in experiment.
\end{abstract}

\maketitle

\textit{Introduction.} -- In recent years, the realization of spin-orbit coupling (SOC) in ultracold atoms \cite{Lin-Nat-471-83-2011,WangPJ-PRL-109-095301-2012,Cheuk-PRL-109-095302-2012,Spielman-Nat-494-49-2013,Huang-NatPhy-12-540-2016,Wu-Sci-354-83-2016} has stimulated intensive studies on the searching of exotic quantum phases brought by SOC. For example, it can lead to unconventional Fermi superfluids\cite{Hu-PRL-107-195304-2011,Chen-PRL-111-235302-2013,Wu-PRL-110-110401-2013,QuC-NatCom-4-2710-2013,Zhang-NatCom-4-2711-2013,Devr-PRL-113-165304-2014,Zheng-PRL-116-120403-2016,WangLL-PRA-95-053628-2017,Sun-PRA-99-043601-2019}, a supersolid stripe phase \cite{Leonard-Nat-543-87-2017,Li-Nat-543-91-2017,Bombin-119-250402-2017}, and diverse magnetic phases\cite{Cole-PRL-109-085302-2012,Radic-PRL-109-085303-2012,Cai-PRA-85-061605-2012,Zhang-NewJPhys-17-073036-2015}. Nevertheless, most of these studies have been focused on the coupling between spin and linear momentum \cite{ZhangJY-2012,Cheuk-2012,Qu-2013,Hamner-2014,Ji-2014,Meng-2016,LiJ-2016}.
More recently by coupling the atomic internal `spin' state to its orbital angular momentum, a new type of SOC which named spin-orbital-angular-momentum coupling (SOAMC), were proposed \cite{Sun-PRA-91-063627-2015,Qu-PRA-91-053630-2015,DeMarco-PRA-91-033630-2015} and firstly realized in spinor BEC \cite{Chen-PRL-121-113204-2018,Chen-PRL-121-250401-2018,Zhang-PRL-122-110402-2019}.

In the SOAMC scheme, a pair of co-propagating high-order Laguerre-Gauassian (LG) lasers \cite{Allen-PRA-45-8185-1992, Babiker-PRL-73-1239-1994, He-PRL-75-826-1995, Marrucci-JP-13-064001-2011} with different orbital angular momenta is generally used to generate an effective coupling between the $z$-component of spin $\vec{\sigma}$ (Pauli representation) and orbital momentum $\vec{L}\equiv \vec{r}\times \vec{p}$ ($\vec{r}$, $\vec{p}$ denotes the position and momentum) in the form of \[H_{\rm SOAMC}\sim\sigma_{z}\cdot L_{z}.\] Such term breaks the translational symmetry while keeps a rotational symmetry, resulting in a {\it discrete} single-particle energy spectrum. This can significantly affect the many-body behaviors of the underlying systems, and give rise to {\it e.g.} the intriguing vortex states in BEC \cite{Zhang-PRL-122-110402-2019, Chen-PRL-121-113204-2018, Wright-PRL-102-030405-2009, Chen-arXiv-1901-02595-2019}. On the other hand,
it would be also interesting and important to explore the unusual superfluid states of a Fermi gas in the presence of SOAMC, which however has never been addressed so far.

\begin{figure}[b]
\centering
\includegraphics[width=0.48\textwidth]{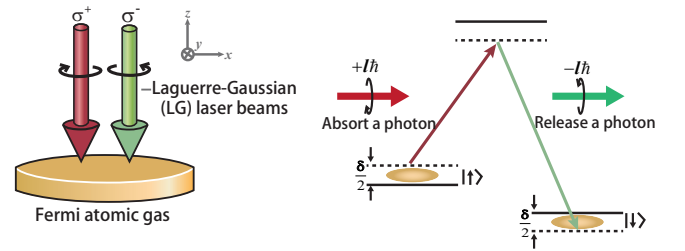}\\
\caption{Schematic of the set-up and energy levels. A pair of LG Raman laser beams with opposite orbital angular momenta $\pm\ell \hbar$ copropagates and generates SOAMC along the $z$-axis in the two-dimensional Fermi gas. $\delta$ is the two-photon detuning.}
\label{Fig:setup}
\end{figure}

In this letter, we investigate the ground state of a two-dimensional (two-component) attractive Fermi gas with SOAMC. By numerically solving the Bogoliubov-de Gennes (BdG) equations self-consistently, we obtain the ground-state phase diagram of the system, which encapsulates rich physics. We show that a family of single-vortex ground states exists in a broad regime of the phase diagram for finite two-photon detuning, in contrast to the conventional case where an external rotation or magnetic field is required. Remarkably, an unprecedented vortex state, which spontaneously breaks the continuous rotational symmetry to a discrete one, is predicted to appear. Such state exhibits a periodical spatial modulation in the angular space, and can be detected in experiment. The underlying mechanism of these vortex states is elucidated to be the unconventional pairing with quantized orbital angular momentum between fermions, which originates from the interplay of SOAMC, two-photon detuning and atom-atom interaction. Our results would open a new avenue to study the unconventional superfluid and vortex physics in Fermi gases with SOAMC.

\textit{The model.} -- We consider a two-component Fermi gas confined in a two-dimensional geometry. As depicted in Fig. \ref{Fig:setup}, the atoms are shined by two copropagating LG laser beams with opposite angular momenta $\pm l\hbar$ to induce an effective SOAMC along $z$-axis via a two-photon Raman process between the two pseudo-spin states (labeled as $\uparrow$ and $\downarrow$). In the transformed basis, the single-particle Hamiltonian of the system can be written as (by setting $\hbar=M=1$) \cite{SM}

\begin{equation}
\begin{split}
 H_{0}  =-\frac{1}{2r}\frac{\partial^2}{\partial r^2}r   & + \frac{L_{z}^2+\ell^2}{2r^2} + \alpha(r) L_{z}\sigma_{z}+ V_{\mathrm{ho}}(r)+\mathcal{L}(r),\\
\mathcal{L}(r)   & = \frac{\delta}{2}\sigma_{z}   + \chi I(r)+ \Omega I(r)\sigma_{x},\\
\end{split}\label{eq01}
\end{equation}
where $L_{z}=-i\partial_{\phi}$ is the $z$ component of the angular momentum operator, which couples to the pseudo-spin $\sigma_{z}$ with inhomogeneous SOAMC strength $\alpha(r)=\ell/r^2$. $\delta$ is the two-photon detuning, the diagonal parameter $\chi$ is the AC Stark light shift and the off-diagonal term $\Omega$ is the effective two-photon Rabi frequency between two pseudo-spin states.
The spatial intensity profile of the LG laser beams is given by
$I(r)=\left(\sqrt{2}\frac{r}{w}\right)^{2|\ell|}[L_{k}^{|\ell|}\left(\frac{2r^2}{w^2}\right)e^{-r^2/w^2}]^2$,
where $w$ characterizes the beam width, and $L^{|\ell|}_{k}$ is the generalized Laguerre polynomials with azimuthal index $\ell$ and the radial index $k$ describing the radial intensity distribution of the LG beams \cite{DeMarco-PRA-91-033630-2015}. $V_{\mathrm{ho}}(r)=\omega_{r}^2r^2/2$ is a harmonic trapping potential with frequency $\omega_{r}$. Without loss of generality, we consider the case with $\ell=4$ and $k=0$.

The single-particle Hamiltonian $H_0$ possesses a rotational symmetry with $[\hat{L}_{z}, H_0]=0$, such that the eigenstates can be labelled by two quantum numbers $(n, m)$ with energies $E_{n,m}$, where $n$ and $m$ are the radial and angular quantum number respectively, giving rise to a discrete single-particle spectrum \cite{SM}. For $\delta=0$, an additional time-reversal (TR) symmetry is hold, i.e. $[\hat{T}, H_0]=0$, where $\hat{T}=i\sigma_y\hat{K}$ with $\hat{K}$ the complex-conjugation operator. This rules out a single vortex ground state and one generally needs a magnetic field or an external rotation to break this symmetry to favor vortex states in a Fermi superfluid. While for $\delta\neq0$, such TR symmetry is absent and we have $E_{n,m}\neq E_{n,-m}$. As shown in below, this would significantly change the pairing between fermions, and a family of exotic vortex states with unusual properties emerges in the ground-state phase diagram.

\begin{figure}[t]
\centering
\includegraphics[width=0.48\textwidth]{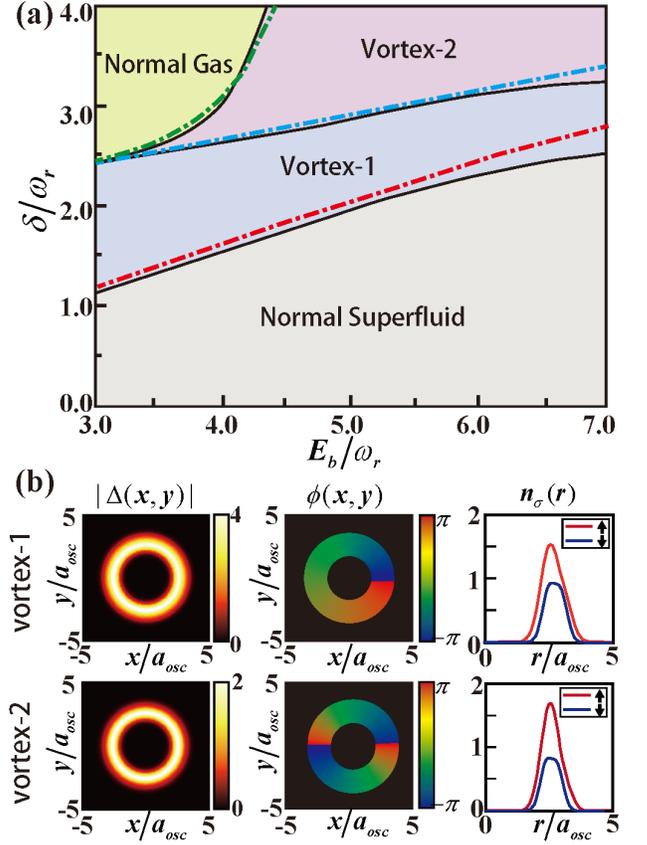}\\
\caption{(Color online) (a) The ground-state phase diagram of the BdG simulations in the $E_b$-$\delta$ plane
for a superfluid Fermi gas with SOAMC. Vortex-$Q$ ($Q=1,2$) denotes a single-vortex state with velocity $Q$. (b) The spatial amplitude $|\Delta(x,y)|$ (left column) and phase $\phi(x,y)$ (middle column) of the order parameter for the vortex phases in (a), and the right column gives the corresponding atomic radial density $n_{\sigma}(r)$ for each vortex. Here we have taken $N_{\mathrm{atom}}=50$, $\ell = 4$, $w=2a_{\mathrm{osc}}$, $\chi/\omega_{r} =-5$, $\Omega/\omega_{r}=1.5$ and $a_{\mathrm{osc}}=\sqrt{1/\omega_{r}}$ being the harmonic oscillator length.}
\label{Fig:phasediagram}
\end{figure}

\textit{Ground-state phase diagram.} -- To reveal the superfluid physics of the system, we consider the $s$-wave contact interaction $H_{\mathrm{int}}=g\int dr \Psi_{\uparrow}^{\dagger}(\mathbf{r})\Psi_{\downarrow}^{\dagger}(\mathbf{r})\Psi_{\downarrow}(\mathbf{r})\Psi_{\uparrow}(\mathbf{r})$ with a bare interaction parameter $g<0$, which is related to the two-body binding energy $E_{b}$ in two dimension via $g=-4\pi/[\ln(1+2E_{c}/E_{b})]$ \cite{Randeria-PRL-62-981-1989}, with $E_{c}$ being the energy cutoff. $\Psi_{\sigma}^{\dagger}(\mathbf{r})$ ($\Psi_{\sigma}(\mathbf{r})$) creates (annihilates) an Fermi atom at position $\mathbf{r}\equiv (r,\phi)$ of spin $\sigma=\uparrow,\downarrow$. Introducing the superfluid order parameter $\Delta(\mathbf{r})=g\langle \Psi_{\downarrow}(\mathbf{r})\Psi_{\uparrow}(\mathbf{r})\rangle$ and applying a Bogoliubov-Valation transformation, the resultant Bogoliubov-de Gennes (BdG) equation is given by \cite{SM}
\begin{equation}
\begin{split}
\begin{bmatrix}
H_{0}-\mu & \Delta(\mathbf{r}) \\
\Delta^{\ast}(\mathbf{r}) & -\sigma_{y}(H_{0}^{\ast}-\mu)\sigma_{y} \\
\end{bmatrix}
\Phi_{\eta}(\mathbf{r})=E_{\eta}\Phi_{\eta}(\mathbf{r})
\end{split},\label{BdG equation}
\end{equation}
with the Nambu representation $\Phi_{\eta}(\mathbf{r})=[u_{\uparrow,\eta}(\mathbf{r}), u_{\downarrow,\eta}(\mathbf{r}), v_{\downarrow,\eta}(\mathbf{r}), -v_{\uparrow,\eta}(\mathbf{r})]^{T}$. $\mu$ is the chemical potential and $E_{\eta}$ is the energy of Bogoliubov quasiparticles labeled by an index $\eta$. In this basis, the order parameter $\Delta(\mathbf{r})$ can be written as
\begin{equation}
\Delta(\mathbf{r})=g\sum_{\eta}[u_{\uparrow,\eta}(\mathbf{r})v^{\ast}_{\downarrow, \eta}(\mathbf{r})f(-E_{\eta})-u_{\downarrow,\eta}(\mathbf{r})v^{\ast}_{\uparrow, \eta}(\mathbf{r})f(E_{\eta})],
\label{Eq:orderparameter}
\end{equation}
where $f(E)=1/[e^{E/k_{B}T}+1]$ is the Fermi-Dirac distribution at a temperature $T$, and the summation is over the quasiparticle state with $E_{\eta}\geq 0$. Self-consistently solving Eq. (\ref{Eq:orderparameter}) and the number equation $N_{\mathrm{atom}}=\sum_{\sigma}\int d\mathbf{r} n_{\sigma}(\mathbf{r})$ with atomic density $n_{\sigma}(r)=\sum_{\eta}[|u_{\sigma,\eta}(r)|^2f(E_{\eta})+|v_{\sigma,\eta}(r)|^2f(-E_{\eta})]$, we can obtain the ground state of the system.

In Fig. \ref{Fig:phasediagram}a, we present the ground-state phase diagram in the $\delta$-$E_{b}$ plane for fixed two-photon Rabi frequency $\Omega/\omega_{r}=1.5$ and $\chi/\omega_{r} =-5$. There are three classes of phases: a normal superfluid (NS) with a real $\Delta\neq0$ (up to a global phase), a normal gas (NG) with $\Delta=0$ and single vortex states. Such a vortex is characterized by a complex $\Delta\neq0$ with a nontrivial phase configuration, i.e. a $2\pi Q$ phase gradient with integer vorticity (winding number) $Q\neq 0$ along a closed path around the vortex (middle column in Fig. \ref{Fig:phasediagram}b). For small $\delta$, the NS is dominant. While for sufficiently large $\delta$, the large energy mismatch between two spin states strongly suppresses the pairing and destroys the superfluid, resulting in a NG phase. In between, a family of vortex states with different vorticity $Q$ exist. In Fig. \ref{Fig:phasediagram}a, we find the regimes for vortex with $Q=1$ and $Q=2$ (labeled as Vortex-1 and Vortex-2), and vortex states with higher $Q$ can be obtained for a larger $\delta$ and $E_b$ (not shown). As $Q$ is quantized, the transition between these phases are of first order.

Considering that an external rotation or magnetic field is generally desired to produce a vortex in a Fermi superfluid without SOAMC, it is quite interesting to see here that the quantum vortex may emerge as a ground state in the presence of SOAMC, suggesting a new mechanism responsible for the forming of vortex in this system. To gain insight, we can write the order parameter as $\Delta=|\Delta(r)|e^{iQ\phi}$ for a single vortex with vorticity $Q$. This implies that pairing with finite orbital angular momentum, if possible, may play a key role in this system. As we will see in below, such exotic pairing state appears naturally from the interplay between SOAMC, two-photon detuning and atom-atom interaction.

\begin{figure}[t]
\centering
\includegraphics[width=0.49\textwidth]{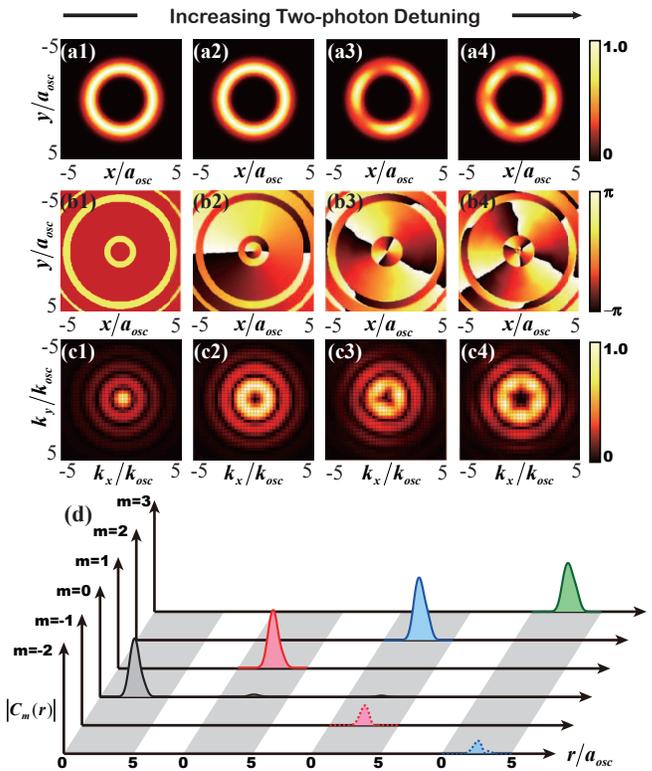}\\
\caption{(Color online) The amplitude (a1-a4) and phase (b1-b4) distributions of the order parameter $\Delta(r)$ in $x$-$y$ plane for different two-photon detuning $\delta/\omega_{r}$= 0, 0.8, 2.2 and 2.8. (c1-c4) are the corresponding momentum distributions $|\Delta(k_{x}, k_{y})|$. (d) The radial distributions $|C_{m}(r)|$ of the $m$-component for order parameters in (a1-a4). Other parameters are $\ell=4$, $w=2a_{\mathrm{osc}}$, $\chi/\omega_{r} = -5$, $E_{b}/\omega_{r}=2$, $N_{\mathrm{atom}}= 50$, $\Omega/\omega_{r}=1$, and the unit of momentum $k_{\mathrm{osc}}$ is defined as $k_{\mathrm{osc}}=\sqrt{2\omega_{r}}$.}
\label{Fig:exoticvortex}
\end{figure}

To grasp the main physics, we notice that the real space density distribution of the superfluid states has a sharp peak around $R = \sqrt{\ell/2} w$ along the radical direction (right column of Fig. \ref{Fig:phasediagram}b) due to the confinement of the red-detuned AC stark potential, it is reasonable to ignore the radial dependence by approximating $r\simeq R$, which leaves an effective one-dimensional model on a ring with the single-particle part $H^{\rm ring}_0=-\partial^2_{\phi}/2R^2-i\tilde{\alpha}\partial_{\phi}\sigma_{z} + (\delta/2)\sigma_{z} + \tilde{\Omega}\sigma_{x}$, where $\tilde{\alpha}=\ell/R^2$, $\tilde{\Omega}=\Omega I(R)$ are the effective SOAMC and Raman coupling respectively. This Hamiltonian can be diagonalized straightforwardly in the basis $\{e^{im\phi}\}$ with $m$ the quantum number of orbital angular momentum. The resultant single-particle energy spectrum has two branches $E_{\pm}(m)=m^2/2R^2\pm\sqrt{\tilde{\Omega}^2+(\tilde{\alpha} m + \delta/2)^2}$. Above results bear some similarity with the case of one-dimensional SOC of linear momentum. However unlike the momentum, here the orbital angular momentum $m$ is quantized, which essentially alters the many-body ground state when the atom-atom interaction is included.

In angular momentum space, the interaction Hamiltonian takes the form of $H^{\rm ring}_{\mathrm{int}}=g\sum_{mm'k}\mathbf{\Psi}^{\dagger}_{m+k\uparrow}\mathbf{\Psi}^{\dagger}_{m'-k\downarrow}\mathbf{\Psi}_{m'\downarrow}\mathbf{\Psi}_{m\uparrow}$, which conserves the total angular momentum.  It is natural to introduce a general pairing field $\Delta_{Q} = g\sum_{m}\left\langle \mathbf{\Psi}_{Q-m\downarrow}\mathbf{\Psi}_{m\uparrow}\right\rangle$ describing the pairing between atoms with angular momenta $m$ and $Q-m$, and write the mean-field Hamiltonian as

\begin{equation}
\begin{split}
H_{\mathrm{\rm MF}}=\frac{1}{2}&\sum_{m}\mathbf{\Phi}_{m,Q}^{\dagger}
\begin{pmatrix}
H_{\mathrm{ring}}(m) & \Delta_{Q}\\
\Delta_{Q} & -\sigma_{y}H_{\mathrm{ring}}^{\ast}(Q-m)\sigma_{y}\\
\end{pmatrix}\mathbf{\Phi}_{m,Q}\\
+ &\sum_{m}\xi_{Q-m} - \frac{|\Delta_{Q}|^2}{g},\\
\end{split}
\label{Eq:meanfield}
\end{equation}
where $\mathbf{\Phi}_{m,Q}=[\mathbf{\Psi}_{m\uparrow}, \mathbf{\Psi}_{m\downarrow}, \mathbf{\Psi}^{\dagger}_{Q-m\downarrow}, -\mathbf{\Psi}^{\dagger}_{Q-m\uparrow}]^{T}$ and $\xi_{m}=m^2/2R^2 - \mu$. Noticing that in the angular space, the above introduced finite orbital angular momentum pairing exactly gives the phase configuration of a vortex state with vorticity $Q$, i.e. $\Delta=\Delta_{Q}e^{iQ\phi}$.

In general, the ground state can be obtained by first minimizing the energy $E^{\rm MF}_Q\equiv\langle H_{\rm MF}\rangle=\sum_{m}\xi_{|Q-m|}+\sum_{m,\nu}\Theta(-E_{m,\nu}^{\eta})E_{m,\nu}^{\eta}-|\Delta_{Q}|^2/g$ for each $Q$, and then further minimizing on $Q$. Here for illustration, we plot the phase boundaries (see the dashed lines in Fig. \ref{Fig:phasediagram}a) by comparing the free energies of different phases with the parameters used in the numerics. The agreement between the numerical simulations and the mean-field results validates the approximation adopted in above, and suggests that the single vortex states originating from the finite orbital angular momentum pairing.

Above we have considered the pairing with a single orbital-angular-momentum component, which is justified by the fermions occupying the lowest energy branch. However, it is also possible that the upper branches may get involved. In that case, pairing with multiple orbital-angular-momentum components may appear, resulting in exotic vortex states with peculiar features.

\textit{Exotic vortex with discrete rotational symmetry.} -- In Fig. \ref{Fig:exoticvortex}, we give the evolution of the ground state as a function of two-photon detuning $\delta$ for a relative weak two-photon Rabi frequency $\Omega/\omega_{r} = 1$. One can see that with the increasing of $\delta$, the NS state with $Q=0$ (Fig. \ref{Fig:exoticvortex}(a1,b1,c1)) first transits into a single vortex state with $Q=1$ (Fig. \ref{Fig:exoticvortex}(a2-c2)), both preserving the rotational symmetry. Note that the momentum distributions of a NS state is always peaked at zero momentum (Fig. \ref{Fig:exoticvortex}(c1)), in sharp comparison to the vortex states with a nearly vanishing zero-momentum contribution and peaked at a finite $k_{r}$ (Fig. \ref{Fig:exoticvortex}(c2-c4)). When further increasing $\delta$, two exotic vortex states become the ground state sequentially (Fig. \ref{Fig:exoticvortex} (a3,a4)). Different from the vortex state governed by a single orbital-angular-momentum pairing, we find the exotic vortex has two different pairing components with angular momentum $Q_1$ and $Q_2$ (referred to Fig. \ref{Fig:exoticvortex}d), by which we can write $\Delta(r,\phi)\sim\sum_{m=Q_1,Q_2}C_m(r)e^{im\phi}$ with $C_m(r)$ the amplitude of $m$-component. Consequently, a spontaneous periodical modulation on the spatial profile of the order parameter $|\Delta|^2\sim C^2_{Q_1}+C^2_{Q_2}+2C_{Q_1}C_{Q_2}\cos(Q_1-Q_2)\phi$ is developed along the azimuthal direction, which breaks the axial symmetry into a discrete $M$-fold symmetry with $M=|Q_1-Q_2|$ in both real and momentum space. Notice that the two components are unequal-weighted, and the winding number of the exotic vortex is determined by the larger one.

\begin{figure}[t]
\centering
\includegraphics[width=0.48\textwidth]{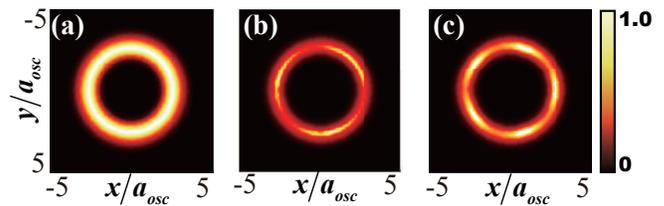}\\
\caption{(Color online) The normalized density distributions of $n(\mathbf{r})$ for different vortex states in Fig. \ref{Fig:exoticvortex}: (a) a single vortex state with rotational symmetry for $\delta/\omega_{r}$= 0.8, (b) an exotic vortex with $\mathcal{C}_3$ symmetry for $\delta/\omega_{r}$= 2.2, and (c) an exotic vortex with $\mathcal{C}_5$ symmetry for $\delta/\omega_{r}$= 2.8.}
\label{Fig:density}
\end{figure}

\textit{Experimental detection.} -- In experiment, the vortex states can be characterized by the total angular momentum $Q$ by measuring the collective shift of the radial quadrupole modes \cite{Chevy-PRL-85-2223-2000,Hodby-PRL-91-090403-2003,Riedl-NewJP-13-035003-2011}. Further, to reveal the exotic vortex phases with discrete rotational symmetry, we turn to their atomic density distributions $n(\mathbf{r})$, which can be directly measured by {\it in-situ} absorption imaging \cite{Bradley-PRL-78-985-1997, Revelle-PRL-117-235301-2016}. In Fig. \ref{Fig:density}, we plot $n(\mathbf{r})=\sum_{\sigma}n_{\sigma}(\mathbf{r})$ for the typical vortex phases discussed above. The reduction from a continuous rotational symmetry to a discrete one can be clearly identified in the density profile, for example a $\mathcal{C}_3$ ($\mathcal{C}_5$) symmetry in Fig. \ref{Fig:density}b (c). This distinguishes the exotic vortex states from the single vortex with rotational symmetry.

\textit{Conclusion.} -- In summary, we have studied the superfluid ground state of a two-dimensional Fermi gas with SOAMC. Due to the finite orbital angular momentum pairing induced by the SOAMC and
two-photon detuning, a family of vortex states with unique features are predicted to be the ground state in a broad regime of the phase diagram. Specifically, an exotic vortex which breaks the rotational symmetry spontaneously may appear. These results reveal the nontrivial physics brought by SOAMC, and open an avenue to search unusual quantum vortex and superfluid phases.

We would like to thank G. Juzeli\={u}nas, Yong Sun, Wenjun Shao and Changan Li for interesting discussions. This work was supported by NSFC under Grants No. 11875195, No. 11474205, foundation of Beijing Education Committees under Grants No. CITTCD201804074, and No. KZ201810028043, and foundation of Zhejiang Province Natural Science under Grant No. LQ20A040002. JL acknowledges support from National Natural Science Foundation of China under Project 11774317. The numerical calculations in this paper have been done on the super-computing system in the Information Technology Center of Westlake University.

\end{document}


\title{Supplemental Material for "Exotic Vortex States with Discrete Rotational Symmetry in Atomic Fermi Gases with Spin-Orbital-Angular-Momentum Coupling"}
\author{Liang-Liang Wang}
\affiliation{Institute for Natural Sciences, Westlake Institute for Advanced Study, Hangzhou, Zhejiang Province, China}
\affiliation{Westlake University, Hangzhou, Zhejiang Province, China}
\author{An-Chun Ji}
\affiliation{Department of Physics,
Capital Normal University, Beijing 100048, China}
\author{Qing Sun}
\email{sunqing@cnu.edu.cn}
\affiliation{Department of Physics,
Capital Normal University, Beijing 100048, China}
\author{Jian Li}
\email{lijian@westlake.edu.cn}
\affiliation{Institute for Natural Sciences, Westlake Institute for Advanced Study, Hangzhou, Zhejiang Province, China}
\affiliation{Westlake University, Hangzhou, Zhejiang Province, China}

\date{{\small \today}}
\maketitle

In this Supplementary Material, we present derivations of the model and the single-particle energy spectrum, and the Bogoliubov-de Gennes equations used in the numerical simulations.

\begin{figure}[t]
\centering
\includegraphics[width=0.85\textwidth]{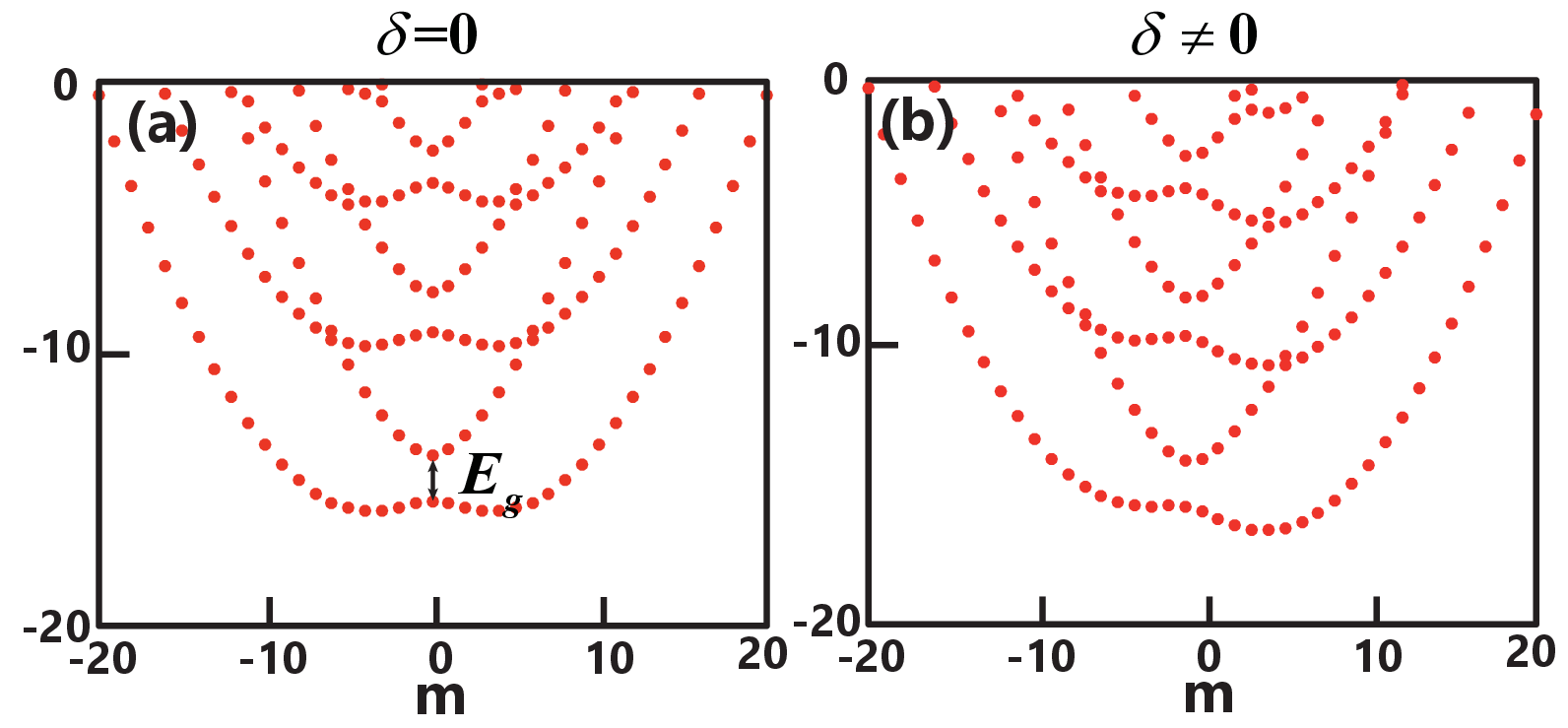}\\
\caption{(Color online) Single particle dispersion relations for different two-photon detuning (a) $\delta/\omega_{r}=0$, (b) $\delta/\omega_{r}=1$. Other parameters are fixed at $\ell = 4$, $w=2a_{\mathrm{osc}}$, $\Omega/\omega_{r}=0.4$, $\chi/\omega_{r}=-5$.}\label{s-fig-01}
\end{figure}

\subsection{The model and single-particle energy spectrum}
We consider $s$-wave interaction Fermi gases in a disk-sharped geometry dressed with two LG laser beams $\sigma^{+}$ and $\sigma^{-}$ propagating collinearly, which carry opposite angular momentum per photon $l^{\pm}=\pm\ell\hbar$, respectively. The integer $\ell$ also defines the phase variation circling the optical axis $z$ with the wavefront phase factor $\exp(\pm i\ell \phi)$. Via a two-photon Raman process, the lasers couple two hyperfine states in the ground electronic mainfold of atom which are labeled as the spin-up $|\uparrow\rangle$ and spin-down $|\downarrow\rangle$ components, respectively\cite{Chen-PRL-121-113204-2018,Zhang-PRL-122-110402-2019}. The single-particle Hamiltonian can be written in polar coordinates $(r, \phi)$ in the following form,
\begin{equation}
H_{0}(r,\phi)=
\begin{pmatrix}
-\frac{1}{2r}\frac{\partial^2}{\partial r^2}r + \frac{L_{z}^2}{2r^2}+ V_{ho}(r)+\frac{\delta}{2}+ \chi_{-} I_{-}(r)+ \chi_{+} I_{+}(r)  & \Omega \sqrt{I_{-}(r)I_{+}(r)}e^{i2l\phi} \\
\Omega\sqrt{I_{-}(r)I_{+}(r)}e^{-i2l\phi} & -\frac{1}{2r}\frac{\partial^2}{\partial r^2}r + \frac{L_{z}^2}{2r^2}+ V_{ho}(r)-\frac{\delta}{2}+ \chi_{-} I_{-}(r)+ \chi_{+} I_{+}(r) \\
\end{pmatrix} \\,\label{Eq:01}
\end{equation}
where  $L_{z}=-i\partial\phi$ is the $z$ component of the angular momentum operator and $I_{+}(r)(I_{-}(r))$ is the intensity of the $\sigma^{+}(\sigma^{-})$ polarized laser field. The diagonal terms in this Hamiltonian containing the parameters $\chi_{\pm}$ are the light shifts and the off-diagonal term $\Omega$ is the effective two-photon Rabi frequency for the coupling between $\uparrow$ and $\downarrow$ states. For simplicity, we assume that the two LG laser beams have identical spatial profile and light intensity with $I(r)=I_{+}(r)=I_{-}(r)$ and $\chi/2 = \chi_{+} = \chi_{-}$, which can be tuned in experiment \cite{Wright-PRL-102-03-405-2009}.  The external harmonic trapping potential is given by $V_{ho}(r) = \omega_{r}^2r^2/2$. Applying a unitary transformation to wave function $\Phi=U\Psi$ with
\begin{equation}
U=
\begin{pmatrix}
e^{-il\phi} & 0 \\
0 & e^{il\phi} \\
\end{pmatrix} \\,\label{Eq:02}
\end{equation}
we can derive the transformed single-particle Hamiltonian $H_0\rightarrow UH_0U^\dag$ given in the main text. Notice that, the above unitary transformation leaves the atom-atom interaction invariant.

To obtain the single-particle energy spectrum, it is convenient to expand the normalized wave functions as
\begin{equation}
\Psi_{\uparrow,\eta}(\mathbf{r})=\sum_{n,m}c_{\uparrow,\eta,n,m}\varphi_{n,m}(\mathbf{r}), \qquad \Psi_{\downarrow,\eta}(\mathbf{r})=\sum_{n,m}d_{\downarrow,\eta,n,m}\varphi_{n,m}(\mathbf{r}),\label{Eq:04}
\end{equation}
with $\varphi_{n,m}(\mathbf{r})=R_{n,m}(r)\Theta_{m}(\phi)$ being the single-particle eigenstates of a quantum harmonic oscillator. Here, $R_{n,m}(r)=\beta^{|m|+1}\sqrt{2n!/(n+|m|)!}e^{-\beta^2r^2/2}r^{|m|}L_{n}^{|m|}(\beta^2r^2)$ and $\Theta_{m}(\phi)=e^{im\phi}/\sqrt{2\pi}$ are the radial and angular part of the wave function respectively, with $m$ denoting the angular momentum quantum number. $\beta=\sqrt{\omega}$ and $L_{n}^{|m|}(x)$ is the associated Laguerre polynomial. Taking use of the orthonormality relations $\int^{\infty}_{0}rdrR_{n,m}(r)R_{n^{'},m}=\delta_{n,n^{'}}$ and $\int^{2\pi}_{0}d\phi\Theta^{\ast}_{m}(\phi)\Theta_{m^{'}}(\phi)=\delta_{m,m^{'}}$ with $\delta$ is the Kronecker delta, we arrive at a Matrix form of the single-particle Hamiltonian which can be diagonalized numerically. In Fig. \ref{s-fig-01}, we show the numerical results of the single-particle spectrum $E_{\eta}$ as a function of $m$. In general, the spectra are {\it discrete} and have multiple branches, with a energy gap $E_g$ between the lowest one and the upper. For $\delta=0$, the spectra are symmetric with respect to $m=0$, reflecting a time reversal symmetry of the system. While for $\delta\neq0$, such time reversal symmetry is broken and the spectra become asymmetric.

\subsection{Bogoliubov-de Gennes equation}
By introducing the superfluid order parameter $\Delta(\mathbf{r})=\mathrm{g}\left\langle\mathbf{\Psi}_{\downarrow}(\mathbf{r})\mathbf{\Psi}_{\uparrow}(\mathbf{r})\right\rangle$ with $\mathbf{r}=(r,\phi)$, the primary Hamiltonian can be diagonalized via a Bogoliubov-Valation transformation and the resultant BdG equation
\begin{equation}
H_{\mathrm{BdG}}(\mathbf{r})\Psi_{\eta}(\mathbf{r})=E_{\eta}\Psi_{\eta}(\mathbf{r})
\end{equation}
is described by an $4\times4$ matrix Hamiltonian
\begin{equation}
H_{\mathrm{BdG}}(\mathbf{r})=
\begin{pmatrix}
H_{0} & \Delta(\mathbf{r})\\
\Delta^{\ast}(\mathbf{r}) & -\sigma_{y}H_{0}^{\ast}\sigma_{y}\\
\end{pmatrix},\label{eq05}
\end{equation}
where $H_{0}$ is the single-particle operator in the primary Hamiltonian. $\Psi_{\eta}(\mathbf{r})\equiv [u_{\uparrow,\eta}(\mathbf{r}),u_{\downarrow,\eta}(\mathbf{r}),v_{\downarrow,\eta}(\mathbf{r}),-v_{\uparrow,\eta}(\mathbf{r})]^{T}$ are the Numbu spinor wave functions corresponding to the quasi-particle excitation energy $E_{\eta}$. The order parameter and the chemical potential can be determined by the self-consistent conditions
\begin{equation}
\Delta(\mathbf{r})= g\sum_{\eta}[u_{\uparrow,\eta}(\mathbf{r})v^{\ast}_{\downarrow,\eta}(\mathbf{r})f(-E_{\eta})-u_{\downarrow,\eta}(\mathbf{r})v^{\ast}_{\uparrow,\eta}(\mathbf{r})f(E_{\eta})]
\end{equation}
and the number equation $N = \int d\mathbf{r}[n_{\uparrow}(\mathbf{r})+n_{\downarrow}(\mathbf{r})]$ with
\begin{equation}
n_{\sigma}(\mathbf{r})=\frac{1}{2}\sum_{\eta}[|u_{\sigma,\eta}(\mathbf{r})|^2f(E_{\eta})+|v_{\sigma,\eta}(\mathbf{r})|^2f(-E_{\eta})]
\end{equation}
being the local density of $\sigma$ fermions. $f(E)=1/[e^{E/K_{B}T}+1]$ is the Fermi distribution function at temperature $T$.

We conveniently expand the normalized wave functions as
\begin{equation}
\begin{split}
u_{\sigma,\eta}(\mathbf{r})=\sum_{n,m}c_{\sigma,\eta,n,m}\varphi_{n,m}(\mathbf{r})\\
v_{\sigma,\eta}(\mathbf{r})=\sum_{n,m}d_{\sigma,\eta,n,m}\varphi_{n,m}(\mathbf{r})\\
\end{split}.
\end{equation}
Here $\varphi_{n,m}(\mathbf{r})=R_{n,m}(r)\Theta_{m}(\phi)$ are the solutions for the single-particle quantum harmonic oscillator problem. We adopt a more general form of the superfluid order parameter
\begin{equation}
\Delta(r,\phi)=\sum_{m}C_{m}(r)e^{im\phi},
\end{equation}
and the solution of the BdG equation then becomes a matrix diagonalization problem,
\begin{equation}
\begin{pmatrix}
K_{\uparrow,n,n'}^{m} & \Omega_{n,n'}^{m} & \Delta_{n,n'}^{m,m'} & 0\\
\Omega_{n,n'}^{m} & K_{\downarrow,n,n}^{m} & 0 & \Delta_{n,n'}^{m,m'}\\
\Delta_{n,n'}^{\ast m,m'} & 0 & -K_{\downarrow,n,n'}^{\ast m} & \Omega_{n,n'}^{m}\\
0 & \Delta_{n,n'}^{\ast m,m'} & \Omega_{n,n'}^{m} & -K_{\uparrow,n,n'}^{\ast m} \\
\end{pmatrix}
\begin{pmatrix}
c_{\uparrow,n',m'}\\
c_{\downarrow,n',m'}\\
d_{\downarrow,n',m'}\\
-d_{\uparrow,n',m'}\\
\end{pmatrix}
=E
\begin{pmatrix}
c_{\uparrow,n,m}\\
c_{\downarrow,n,m}\\
d_{\downarrow,n,m}\\
-d_{\uparrow,n,m}\\
\end{pmatrix},
\end{equation}
where the matrix elements
\begin{equation}
\begin{split}
&K_{\uparrow,n,n'}^{m}= [(2n+|m+\ell|+1)\hbar\omega-\mu+\frac{\delta}{2}]\delta_{n,n'}+2\chi\int^{\infty}_{0}rdrR_{n,m}(r)I(r)R_{n',m}(r),\\
&K_{\downarrow,n,n'}^{m}= [(2n+|m-\ell|+1)\hbar\omega-\mu-\frac{\delta}{2}]\delta_{n,n'}+2\chi\int^{\infty}_{0}rdrR_{n,m}(r)I(r)R_{n',m}(r),\\
&K_{\uparrow,n,n'}^{\ast m}= [(2n+|m-\ell|+1)\hbar\omega-\mu+\frac{\delta}{2}]\delta_{n,n'}+2\chi\int^{\infty}_{0}rdrR_{n,m}(r)I(r)R_{n',m}(r),\\
&K_{\downarrow,n,n'}^{\ast m}= [(2n+|m+\ell|+1)\hbar\omega-\mu-\frac{\delta}{2}]\delta_{n,n'}+2\chi\int^{\infty}_{0}rdrR_{n,m}(r)I(r)R_{n',m}(r),\\
&\Omega_{n,n'}^{m}= \Omega\int^{\infty}_{0}rdrR_{n,m}(r)I(r)R_{n',m}(r),\\
&\Delta_{n,n'}^{m,m'}= \sum_{m_{0}}\delta_{m,m_{0}+m'}\int^{\infty}_{0}rdrR_{n,m}(r)C_{m_{0}}(r)R_{n',m'}(r),\\
&\Delta_{n,n'}^{\ast m,m'}= \sum_{m_{0}}\delta_{m_{0}+m,m'}\int^{\infty}_{0}rdrR_{n,m}(r)C_{m_{0}}(r)R_{n',m'}(r).\\
\end{split}
\end{equation}
The same procedure also reduces the order parameter equation to
\begin{equation}
\Delta(\mathbf{r})=\frac{\mathrm{g}}{2\pi}\sum_{n,n',m,m',\eta}[c^{\eta}_{\uparrow n,m}d^{\ast\eta}_{\downarrow n',m'}R_{n,m}(r)R_{n',m'}(r)e^{i(m'-m)\phi}f(-E_{\eta})-c^{\eta}_{\downarrow n,m}d^{\ast\eta}_{\uparrow n',m'}R_{n,m}(r)R_{n',m'}(r)e^{i(m'-m)\phi}f(E_{\eta})],
\end{equation}
and the local-density equation to
\begin{equation}
n_{\sigma}(\mathbf{r})=\frac{1}{2\pi}\sum_{n,n',m,m',\eta}[c^{\eta}_{\sigma n,m}c^{\ast\eta}_{\sigma n',m'}R_{n,m}(r)R_{n',m'}(r)e^{i(m'-m)\phi}f(E_{\eta})+d^{\eta}_{\sigma n,m}d^{\ast\eta}_{\sigma n',m'}R_{n,m}(r)R_{n',m'}(r)e^{i(m'-m)\phi}f(-E_{\eta})].
\end{equation}
We recall that the sums are only over the quasi-particle states with $E_{\eta}\geq0$. Using the orthonormality relations, we also obtain the total number of $\sigma$ fermions as
\begin{equation}
N_{\sigma}=\sum_{m,n,\eta}[|c^{\eta}_{\sigma n,m}|^2f(E_{\eta})+|d^{\eta}_{\sigma n,m}|^2f(E_{-\eta})].
\end{equation}
Numerically we have to truncate the summation over energy levels $\eta$. This is done by introducing a high energy cut-off $E_{c}$.